\documentclass[11pt,a4paper,reqno,intlimits,sumlimits]{amsart}

\usepackage{amssymb}
\usepackage{chicago}
\usepackage{color}
\usepackage[mathscr]{euscript}
\usepackage{xspace}
\usepackage{epsfig,rotating}
\usepackage{enumerate}

\DeclareMathAlphabet{\mathpzc}{OT1}{pzc}{m}{it}

\usepackage[bookmarksopen,pdfstartview=FitH]{hyperref}
\hypersetup{colorlinks,%
            citecolor=black,%
            filecolor=black,%
            linkcolor=black,%
            urlcolor=black}%

\begin{document}

\frenchspacing

\theoremstyle{plain}
\newtheorem{theorem}{Theorem}[section]
\newtheorem{lemma}[theorem]{Lemma}
\newtheorem{proposition}[theorem]{Proposition}
\newtheorem{corollary}[theorem]{Corollary}

\theoremstyle{definition}
\newtheorem{remark}[theorem]{Remark}
\newtheorem{definition}[theorem]{Definition}
\newtheorem{assumption}{Assumption}
\newtheorem*{assuL}{Assumption ($\mathbb{L}$)}
\newtheorem*{assuAC}{Assumption ($\mathbb{AC}$)}
\newtheorem*{assuEM}{Assumption ($\mathbb{EM}$)}
\renewcommand{\theequation}{\thesection.\arabic{equation}}
\numberwithin{equation}{section}

\renewcommand{\thetable}{\thesection.\arabic{table}}
\numberwithin{table}{section}

\renewcommand{\thefigure}{\thesection.\arabic{figure}}
\numberwithin{figure}{section}

\newcommand{\Law}{\ensuremath{\mathop{\mathrm{Law}}}}
\newcommand{\loc}{{\mathrm{loc}}}
\newcommand{\Log}{\ensuremath{\mathop{\mathcal{L}\mathrm{og}}}}

\let\SETMINUS\setminus
\renewcommand{\setminus}{\backslash}

\def\stackrelboth#1#2#3{\mathrel{\mathop{#2}\limits^{#1}_{#3}}}

\newcommand{\prozess}[1][L]{{\ensuremath{#1=(#1_t)_{0\le t\le T_*}}}\xspace}
\newcommand{\prazess}[1][L]{{\ensuremath{#1=(#1_t)_{0\le t\le T_*}}}\xspace}
\newcommand{\pt}[1][]{{\ensuremath{\P_{T_{#1}}}}\xspace}
\newcommand{\ts}[1][]{\ensuremath{T_{#1}}\xspace}
\def\P{\ensuremath{\mathrm{I\kern-.2em P}}}
\def\E{\mathrm{I\kern-.2em E}}

\def\bF{\mathbf{F}}
\def\F{\ensuremath{\mathcal{F}}}
\def\R{\ensuremath{\mathbb{R}}}
\def\C{\ensuremath{\mathbb{C}}}
\def\bt{\ensuremath{\mathbf{T}}}

\def\Rmz{\R\setminus\{0\}}
\def\Rdmz{\R^d\setminus\{0\}}
\def\Rnmz{\R^n\setminus\{0\}}

\def\Rp{\mathbb{R}_{\geqslant0}}

\def\lev{L\'{e}vy\xspace}
\def\lib{LIBOR\xspace}
\def\lk{L\'{e}vy--Khintchine\xspace}
\def\smmg{semimartingale\xspace}
\def\smmgs{semimartingales\xspace}
\def\mg{martingale\xspace}
\def\tih{time-inhomoge\-neous\xspace}

\def\eqlaw{\ensuremath{\stackrel{\mathrrefersm{d}}{=}}}

\def\ud{\ensuremath{\mathrm{d}}}
\def\dt{\ud t}
\def\ds{\ud s}
\def\dx{\ud x}
\def\dy{\ud y}
\def\dsdx{\ensuremath{(\ud s, \ud x)}}
\def\dtdx{\ensuremath{(\ud t, \ud x)}}

\def\intrr{\ensuremath{\int_{\R}}}

\def\MM{\ensuremath{\mathscr{M}}}
\def\ME{\mathbb{M}}

\def\EM{\ensuremath{(\mathbb{EM})}\xspace}
\def\ES{\ensuremath{(\mathbb{ES})}\xspace}
\def\AC{\ensuremath{(\mathbb{AC})}\xspace}
\def\LL{\ensuremath{(\mathbb{L})}\xspace}

\def\ott{{0\leq t\leq T_*}}

\def\e{\mathrm{e}}
\def\eps{\epsilon}

\def\half{\frac{1}{2}}
\def\LibT{L(t,T_i)}
\def\MeaT{\P_{T_{i+1}}}
\def\volT{\lambda(s,T_i)}
\def\vol2T{\lambda^2(s,T_i)}
\def\LevT{H_s^{T_{i+1}}}

\title[Picard Approximation of SDE and LIBOR models]
      {Picard approximation of stochastic differential equations\\
       and application to LIBOR models}

\author[A. Papapantoleon]{Antonis Papapantoleon}
\author[D. Skovmand]{David Skovmand}

\address{Institute of Mathematics, TU Berlin, Stra\ss e des 17. Juni 136,
         10623 Berlin, Germany \& Quantitative Products Laboratory,
         Deutsche Bank AG, Alexanderstra\ss e 5, 10178 Berlin, Germany}
\email{papapan@math.tu-berlin.de}

\address{Aarhus School of Business, Aarhus University, Fuglesangs All\'e 4,
         8210 Aarhus V, Denmark}
\email{davids@asb.dk}

\thanks{We would like to thank Friedrich Hubalek and Peter Tankov for
        interesting discussions during the work on these topics. We are also
        grateful to two anonymous referees for their careful reading and
        constructive comments.}

\keywords{LIBOR models, L\'evy processes, Picard approximation, drift expansion,
          parallel computing}
\subjclass[2000]{91G30, 91G60, 60G51 (2010 MSC)}

\date{}
\maketitle
\pagestyle{myheadings}

\begin{abstract}
The aim of this work is to provide fast and accurate approximation schemes for
the Monte Carlo pricing of derivatives in \lib market models. Standard methods
can be applied to solve the stochastic differential equations of the successive
\lib rates but the methods are generally slow. Our contribution is twofold.
Firstly, we propose an alternative approximation scheme based on Picard
iterations. This approach is similar in accuracy to the Euler discretization,
but with the feature that each rate is evolved independently of the other rates
in the term structure. This enables simultaneous calculation of derivative
prices of different maturities using parallel computing. Secondly, the product
terms occurring in the drift of a LIBOR market model driven by a jump process
grow exponentially as a function of the number of rates, quickly rendering the
model intractable. We reduce this growth from exponential to quadratic using
truncated expansions of the product terms. We include numerical illustrations
of the accuracy and speed of our method pricing caplets, swaptions and forward
rate agreements.
\end{abstract}

\section{Introduction}
\label{intro}

The \lib market model (LMM) has become a standard model for the pricing of
interest rate derivatives in recent years. The main advantage of this model in
comparison to other approaches is that the evolution of discretely compounded,
market-observable forward rates is modeled directly and not deduced from the
evolution of unobservable factors. Moreover, the log-normal \lib model is
consistent with the market practice of pricing caps according to Black's formula
(cf. \citeNP{Black76}). However, despite its apparent popularity, the \lib
market model has certain well-known pitfalls.

On the one hand, the log-normal \lib model is driven by a Brownian motion, hence
it cannot be calibrated adequately to the observed market data. An interest rate
model is typically calibrated to the implied volatility surface from the cap
market and the correlation structure of at-the-money swaptions. Several
extensions of the \lib model have been proposed in the literature using
jump-diffusions, \lev processes or general semimartingales as the driving motion
(cf. e.g. \citeNP{GlassermanKou03}, Eberlein and \"Ozkan
\citeyearNP{EberleinOezkan05},
\citeNP{Jamshidian99}), or incorporating stochastic volatility effects (cf. e.g.
\citeNP{AndersenBrothertonRatcliffe05}).

On the other hand, the dynamics of LIBOR rates are not tractable under forward
measures due to the random terms that enter the dynamics of rates during the
construction of the model. In particular, when the driving process has
continuous paths the dynamics are tractable under their corresponding forward
measure, but not under any other forward measure. When the driving process is a
general semimartingale, then the dynamics of \lib rates are not even tractable
under their very own forward measure. Consequently:
\begin{enumerate}
\item if the driving process is a continuous \smmg caplets can be
      priced in ``closed form'', but not swaptions or other multi-LIBOR
      derivatives;
\item if the driving process is a general \smmg, then even caplets cannot be
      priced in closed form.
\end{enumerate}

The standard remedy to this problem is the so-called ``frozen drift''
approximation; it was first proposed by \shortciteNP{BraceGatarekMusiela97} for
the pricing of swaptions and has been used by several authors ever since.
\shortciteANP{BraceDunBarton01} \citeyear{BraceDunBarton01},
\shortciteN{DunBartonSchloegl01} and
\citeN{Schloegl02} argue that freezing the drift is justified, since the
deviation from the original equation is small in several measures.

Although the frozen drift approximation is the simplest and most popular
solution, it is well-known that it does not yield acceptable results, especially
for exotic derivatives and longer horizons. Therefore, several other
approximations have been developed in the literature. In one line of research
\citeN{DanilukGatarek05} and \shortciteN{KurbanmuradovSabelfeldSchoenmakers} are
looking for the best lognormal approximation of the forward LIBOR dynamics; cf.
also \citeN{Schoenmakers05}. Other authors have been using linear interpolations
and predictor-corrector Monte Carlo methods to get a more accurate approximation
of the drift term (cf. e.g. \shortciteNP{HunterJaeckelJoshi01} and
Glasserman and Zhao \citeyearNP{GlassermanZhao00}). We refer the reader to
\citeN{JoshiStacey08} for a
detailed overview of that literature, some new approximation schemes and
numerical experiments.

Although most of this literature focuses on the lognormal LIBOR market model,
\citeANP{GlassermanMerener03} (\citeyearNP{GlassermanMerener03},
\citeyearNP{GlassermanMerener03b}) have developed approximation schemes for the
pricing of caps and swaptions in jump-diffusion \lib market models.

In this article we develop a general method for the approximation of the random
terms that enter the drift of LIBOR models that is suitable for parallel
computing. In particular, using Picard iterations we develop generic
approximation schemes that decouple the dependence between \lib rates. Therefore
individual rates in the tenor can be evolved independently in a Monte Carlo
simulation. In addition, we treat a problem specific to LIBOR models with jumps;
namely that the complexity of the drift term grows exponentially in the number
of tenor dates. We expand and truncate the drift term, which yields a highly
accurate approximation, while the gain in computational speed is immense. We
illustrate the accuracy and efficiency of our method in an example where LIBOR
rates are driven by a normal inverse Gaussian process.

The method we develop is universal and can be applied to any \lib model driven
by a general \smmg. However, we focus on the \lev \lib model as a characteristic
example of a \lib model driven by a general semimartingale.

The article is structured as follows: in section \ref{PIIAC} we review \tih
\lev process, and in section \ref{LevyLIBOR} we revisit the \lev \lib model and
explain in detail the computational problems. In section \ref{Picard-section}
we derive the Picard approximation scheme and the drift expansions. In section
\ref{derivatives} we briefly describe the main derivatives on \lib rates.
Finally, section \ref{numerics} contains a numerical illustration.

\section{L\'evy processes}
\label{PIIAC}

Let ($\Omega, \F, \bF, \P$) be a complete stochastic basis, where $\F=\F_{T_*}$
and the filtration $\bF=(\F_t)_{t\in[0,T_*]}$ satisfies the usual conditions;
we assume that $T_*\in\Rp$ is a finite time horizon. The driving process
\prozess[H] is a \emph{process} with \emph{independent increments} and
\emph{absolutely continuous} characteristics; this is also called a \emph{\tih
\lev process}. That is, $H$ is an adapted, c\`{a}dl\`{a}g, real-valued
stochastic
process with independent increments, starting from zero, where the law of $H_t$,
$t\in[0,T_*]$, is described by the characteristic function
\begin{align}\label{LK}
\E\!\left[\e^{iuH_{t}}\right]
 = \exp\left(\int_{0}^{t}\Big[ ib_su - \frac{c_s}{2}u^{2}
  + \int_{\R}(\e^{iux}-1-iux)F_s(\ud x)\Big]\ud s\right);
\end{align}
here $b_t\in\R$, $c_t\in\Rp$ and $F_t$ is a \lev measure, i.e. satisfies
$F_t(\{0\})=0$
and $\int_{\R}(1\wedge|x|^2)F_t(\ud x)<\infty$, for all $t\in[0,T_*]$. In
addition,
the process $H$ satisfies Assumptions ($\mathbb{AC}$) and ($\mathbb{EM}$) given
below.

\begin{assuAC}
The triplets ($b_t,c_t,F_t$) satisfy
\begin{eqnarray}
\int_{0}^{T_*}\bigg( |b_t| + c_t
  + \int_{\R}(1\wedge|x|^2)F_t(\ud x) \bigg)\ud t <\infty.
\end{eqnarray}
\end{assuAC}

\begin{assuEM}
There exist constants $M, \varepsilon>0$ such that for every
$u\in[-(1+\varepsilon)M,(1+\varepsilon)M]=:\ME$
\begin{equation}\label{eq:Int}
    \int_0^{T_*}\int_{\{|x|>1\}}\e^{ux} F_t(\ud x)\ud t<\infty.
\end{equation}
Moreover, without loss of generality, we assume that
$\int_{\{|x|>1\}}\e^{ux}F_t(\ud x)<\infty$ for all $t\in[0,T_*]$ and $u\in\ME$.
\end{assuEM}

These assumptions render the process \prozess[H] a \emph{special}
semimartingale, therefore it has the canonical decomposition (cf.
Jacod and Shiryaev
\citeyearNP[II.2.38]{JacodShiryaev03}, and \shortciteNP{EberleinJacodRaible05})
\begin{align}\label{canonical}
 H = \int_{0}^{\cdot}b_s\ud s
       + \int_{0}^{\cdot}\sqrt{c_s} \ud W_{s}
       + \int_{0}^{\cdot}\int_{\R} x(\mu^{H}-\nu)(\ud s,\ud x),
\end{align}
where $\mu^H$ is the random measure of jumps of the process $H$, $\nu$ is the
$\P$-compensator of $\mu^H$, and \prazess[W] is a $\P$-standard Brownian motion.
The \emph{triplet of predictable characteristics} of $H$ with respect to the
measure $\P$, $\mathbb T(H|\P)=(B,C,\nu)$, is
\begin{eqnarray}\label{ch4:triplet}
 B = \int_{0}^{\cdot}b_s\ud s, \qquad
 C = \int_0^\cdot c_s\ud s, \qquad
 \nu([0,\cdot] \times A)=\int_0^\cdot\int_A F_s(\ud x)\ud s,
\end{eqnarray}
where $A\in\mathcal{B}(\R)$; the triplet ($b,c,F$) represents the \emph{local
characteristics} of $H$. In addition, the triplet of predictable characteristics
($B,C,\nu$) determines the distribution of $H$, as the \lk formula \eqref{LK}
obviously dictates.

We denote by $\kappa_s$ the \emph{cumulant generating function} associated to
the infinitely divisible distribution with \lev triplet ($b_s,c_s,F_s$), i.e.
for $z\in\ME$ and $s\in[0,T_*]$
\begin{align}\label{cumulant}
\kappa_s(z) := b_sz+\frac{c_s}{2}z^{2}
             + \int_{\R}(\e^{zx}-1-zx)F_s(\ud x).
\end{align}
Using Assumption \EM we can extend $\kappa_s$ to the complex domain $\C$, for
$z\in\C$ with $\Re z\in\ME$, and the characteristic function of $H_t$ can be
written as
\begin{align}\label{char-fun}
\E\!\left[\e^{iuH_{t}}\right] = \exp\bigg(\int_{0}^{t} \kappa_s(iu)\ud s\bigg).
\end{align}
If $H$ is a \lev process, i.e. time-homogeneous, then ($b_s,c_s,F_s$) -- and
thus
also $\kappa_s$ -- do not depend on $s$. In that case, $\kappa$ equals the
cumulant
(log-moment) generating function of $H_1$.

\section{The L\'evy LIBOR model}
\label{LevyLIBOR}

\subsection{Model description}

The \lev LIBOR model was developed by Eberlein and \"Ozkan
\citeyear{EberleinOezkan05} following
the seminal articles on LIBOR market models driven by Brownian motion by
\shortciteN{SandmannSondermannMiltersen95},
\shortciteANP{MiltersenSandmannSondermann97} \citeyear{MiltersenSandmannSondermann97} and
\shortciteN{BraceGatarekMusiela97}; see also \citeN{GlassermanKou03} and
\citeN{Jamshidian99} for LIBOR market models driven by jump processes and
general semimartingales respectively. The \lev \lib model is a
\textit{market model} where the forward LIBOR rate is modeled directly
and is driven by a \tih \lev process.

Let $0=T_0<T_{1}<\cdots<T_{N}<T_{N+1}=T_*$ denote a discrete tenor
structure where $\delta_i=T_{i+1}-T_{i}$, $i\in\{0,1,\dots,N\}$. Consider a
complete stochastic basis $(\Omega, \F,\mathbf{F},\P_{T_*})$ and a \tih \lev
process \prozess[H] satisfying Assumptions $(\mathbb{AC})$ and $(\mathbb{EM})$.
The process $H$ has predictable characteristics ($0,C,\nu^{T_*}$) or local
characteristics $(0,c,F^{T_*})$, and its canonical decomposition is
\begin{align}\label{canon-LIBOR}
H = \int_0^\cdot \sqrt{c_s}\ud W_s^{T_*}
  + \int_0^\cdot\int_{\R}x(\mu^H-\nu^{T_*})\dsdx,
\end{align}
where $W^{T_*}$ is a $\P_{T_*}$-standard Brownian motion, $\mu^H$ is the random
measure associated with the jumps of $H$ and $\nu^{T_*}$ is the
$\P_{T_*}$-compensator of $\mu^H$. We further assume that the following
conditions are in force.
\begin{description}
\item[(LR1)] For any maturity $T_{i}$ there exists a bounded, continuous,
             deterministic function $\lambda(\cdot,T_{i}):[0,T_i]\rightarrow
\R$,
             which represents the volatility of the forward LIBOR rate process
             $L(\cdot, T_{i})$. Moreover,
\begin{align*}
\sum_{i=1}^N \big|\lambda(s,T_i)\big|\leq M,
\end{align*}
             for all $s\in[0,T_*]$, where $M$ is the constant from Assumption
             ($\mathbb{EM}$), and $\lambda(s,T_i)=0$ for all $s>T_i$.
\item[(LR2)] The initial term structure $B(0,T_i)$, $1\leq i\leq N+1$, is
strictly
             positive and strictly decreasing. Consequently, the initial term
             structure of forward LIBOR rates is given, for $1\leq i\leq N$, by
\begin{align}\label{i-val}
L(0,T_i)=\frac{1}{\delta_i}\left(\frac{B(0,T_i)}{B(0,T_i+\delta_i)}-1\right)>0.
\end{align}
\end{description}

The construction of the \lev \lib model starts by postulating that the dynamics
of the forward LIBOR rate with the longest maturity $L(\cdot,T_N)$ is driven by
the \tih \lev process $H$ and evolve as a martingale under the terminal forward
measure $\P_{T_*}$. Then, the dynamics of the LIBOR rates for the preceding
maturities are constructed by backward induction; they are driven by the same
process $H$ and evolve as martingales under their corresponding forward
measures.

Let us denote by $\MeaT$ the forward measure associated to the settlement date
$T_{i+1}$, $i\in\{0,\dots,N\}$. The dynamics of the forward LIBOR rate
$L(\cdot,T_i)$, for an arbitrary $T_i$, is given by
\begin{align}\label{LIBOR-dyn}
\LibT = L(0,T_i)
 \exp\left(\int_0^t b^L(s,T_i)\ud s+\int_0^t \volT\ud\LevT\right),
\end{align}
where $H^{T_{i+1}}$ is a special \textit{semimartingale} with canonical
decomposition
\begin{align}\label{LIBOR-Levy}
H_t^{T_{i+1}}
      = \int_0^t \sqrt{c_s}\ud W_s^{T_{i+1}}
      + \int_0^t\int_{\R}x(\mu^H-\nu^{T_{i+1}})\dsdx.
\end{align}
Here $W^{T_{i+1}}$ is a $\P_{T_{i+1}}$-standard Brownian motion and
$\nu^{T_{i+1}}$ is the $\P_{T_{i+1}}$-compensator of $\mu^H$. The dynamics of
an arbitrary LIBOR rate again evolves as a martingale under its corresponding
forward measure; therefore, we specify the drift term of the forward LIBOR
process $L(\cdot,T_i)$ as
\begin{align}\label{LIBOR-drift-term}
b^L(s,T_i)
 & = -\half \vol2T c_s
     - \int_{\R} \big(\e^{\volT x}-1-\volT x\big)F_s^{T_{i+1}}(\ud x).
\end{align}

The forward measure $\MeaT$, which is defined on
$(\Omega,\F,(\F_t)_{0\leq t\leq T_{i+1}})$, is related to the terminal forward
measure $\P_{T_*}$ via
\begin{align}\label{TF-libor}
\frac{\ud\MeaT}{\ud\P_{T_*}}
 = \prod_{l=i+1}^{N}\frac{1+\delta_l L(T_{i+1},T_l)}{1+\delta_l L(0,T_l)}
 = \frac{B(0,T_*)}{B(0,T_{i+1})}\prod_{l=i+1}^{N} \left(1+\delta_l
L(T_{i+1},T_l)\right).
\end{align}
The $\MeaT$-Brownian motion $W^{T_{i+1}}$ is related to the $\P_{T_*}$-Brownian
motion via
\begin{align}\label{Levy-LIBOR-Brownian}
W_t^{T_{i+1}}
 & = W_t^{T_{i+2}} - \int_0^t \alpha(s,T_{i+1})\sqrt{c_s}\ud s
   = \dots\nonumber\\
 & = W_t^{T_*} - \int_0^t \left(\sum_{l=i+1}^{N}\alpha(s,T_l)\right)\sqrt{c_s}
\ud s,
\end{align}
where
\begin{align}\label{Levy-LIBOR-alpha}
\alpha(t,T_l)
 = \frac{\delta_l L(t-,T_l)}{1+\delta_l L(t-,T_l)}\lambda(t,T_l).
\end{align}
The $\MeaT$-compensator of $\mu^H$, $\nu^{T_{i+1}}$, is related to the
$\P_{T_*}$-compensator of $\mu^H$ via
\begin{align}\label{Levy-LIBOR-compensator}
\nu^{T_{i+1}}\dsdx
 &= \beta(s,x,T_{i+1})\nu^{T_{i+2}}\dsdx
  = \dots\nonumber\\
 &= \left(\prod_{l=i+1}^{N}\beta(s,x,T_l)\right)\nu^{T_*}\dsdx,
\end{align}
where
\begin{align}\label{Levy-LIBOR-beta}
\beta(t,x,T_l,)
 = \frac{\delta_l L(t-,T_l)}{1+\delta_l
L(t-,T_l)}\Big(\e^{\lambda(t,T_l)x}-1\Big) +1.
\end{align}

\begin{remark}\label{LIBOR-conn}
Notice that the process $H^{T_{i+1}}$, driving the forward LIBOR rate
$L(\cdot,T_i)$,
and $H=H^{T_*}$ have the same \emph{martingale} part and differ only in the
\emph{finite variation} part (drift). An application of Girsanov's theorem for
semimartingales yields that the $\P_{T_{i+1}}$-finite variation part of $H$ is
\begin{align*}
\int_0^{\cdot} c_s\sum_{l=i+1}^{N}\alpha(s,T_l) \ud s
  + \int_0^{\cdot}\intrr
x\left(\prod_{l=i+1}^{N}\beta(s,x,T_l)-1\right)\nu^{T_*}\dsdx.
\end{align*}
\end{remark}

\subsection{Option pricing and computational problems}

The main scope of a market model for interest rate derivatives is to adequately
describe the dynamics of interest rates as they are reflected in prices of
derivatives. Hence, a good market model should be easily calibrated to option
prices of liquid derivatives, i.e. caps and at-the-money swaptions. Calibration
requires the fast computation of option prices; either in closed-form or using
semi-analytical methods (e.g. Fourier transform methods).

However, herein lies a major pitfall of the \lev \lib model: the process
$H^{T_{i+1}}$ driving the dynamics of $L(\cdot,T_i)$ is \emph{not} a \lev
process under $\P_{T_{i+1}}$, or any other forward measure. One just has
to observe that the compensator $\nu^{T_{i+1}}$ is random and not
deterministic. Therefore, the characteristic function of the random
variable $H_t^{T_{i+1}}$ is not available and Fourier methods cannot be
used for option pricing. In other words, on top of the well-known problems
of LMMs in pricing swaptions and other multi-\lib products, when the
driving process has jumps even caplets \emph{cannot} be priced in
closed or semi-analytic form.

A remedy has been proposed in \citeN{EberleinOezkan05} and further refined
by \citeN{Kluge05}. They propose to ``freeze'' the random terms in the
compensator, i.e. to replace them by their deterministic initial values.
The approximate process is then a \lev process and Fourier methods for
option pricing can be applied. This method is equivalent to the ``frozen
drift'' approximation, which does not however yield acceptable results.

\subsection{Terminal measure dynamics}
\label{dynamics}

Once closed-form or semi-analytical methods are not available for option
pricing, a Monte Carlo simulation is the next alternative. In this section
we derive the dynamics of \lib rates under the terminal measure. This is
the appropriate measure for simulations in LMMs.

Starting with the dynamics of the LIBOR rate $L(\cdot,\ts[i])$ under the
forward martingale measure \pt[i+1], and using the connection between the
forward and terminal martingale measures (cf. eqs.
\eqref{Levy-LIBOR-Brownian}--\eqref{Levy-LIBOR-beta} and Remark
\ref{LIBOR-conn}), we have that the dynamics of the LIBOR rate $L(\cdot,\ts[i])$
under the terminal measure is given by
\begin{align}\label{LIBOR-dyn-PT}
\LibT = L(0,T_i)
 \exp\left( \int_0^t b(s,T_i)\ud s+\int_0^t \volT\ud H_s \right),
\end{align}
where \prozess[H] is the $\P_{T_*}$-\tih \lev process driving the LIBOR rates,
cf. \eqref{canon-LIBOR}. The drift term $b(\cdot,\ts[i])$ has the form
\begin{align}\label{LIBOR-drift-PT}
b(s,\ts[i])
 & = -\half \vol2T c_s
     - c_s\volT \sum_{l=i+1}^{N}\frac{\delta_l L(s-,T_l)}{1+\delta_l
L(s-,T_l)}\lambda(s,T_l)
     \nonumber\\&\quad
     - \int_{\R}\left(\Big(\e^{\volT x}-1\Big)\prod_{l=i+1}^{N}\beta(s,x,T_l)
                      -\volT x\right)F_s^{T_*}(\ud x),
\end{align}
where $\beta(s,x,T_l)$ is given by \eqref{Levy-LIBOR-beta}. Note that the
drift term of \eqref{LIBOR-dyn-PT} is random, therefore the log-\lib is a
general semimartingale, and not a \lev process. Of course, $L(\cdot,\ts[i])$
is not a $\P_{T_*}$-martingale, unless $i=N$ (where we use the conventions
$\sum_{l=1}^0=0$ and $\prod_{l=1}^0=1$).

The equation for the dynamics under the terminal measure contains the
next numerical problem, well-known for LMMs. The drift $b(\cdot,\ts[i])$
depends on all \emph{subsequent} \lib rates in the tenor, yielding a
dependence structure that has the form of a triangular matrix;
cf. Table \ref{LIBOR-matrix}. In other words, in order to simulate any
rate one has to start by simulating the last rate, save the path and
proceed iteratively. This means that simulations are very slow, while
the burden on the random access memory (RAM) is also significant.

{\footnotesize\renewcommand{\arraystretch}{1.2}
\begin{table}[h]
 \begin{center}
  \begin{tabular}{cccccccc}
\hline
$L(t,T_1)$      & $L(t,T_2)$      & \dots  & $L(t,T_k)$     & \dots  &
$L(t,\ts[N-2])$ & $L(t,\ts[N-1])$ & $L(t,T_N)$\\
\hline
$L(t,T_N)$      & $L(t,T_N)$      & \dots  & $L(t,T_N)$     & \dots  &
$L(t,T_N)$      & $L(t,T_N)$      & \\
$L(t,\ts[N-1])$ & $L(t,\ts[N-1])$ & \dots  & $L(t,T_{N-1})$ & \dots  &
$L(t,\ts[N-1])$ &&\\
$L(t,\ts[N-2])$ & $L(t,\ts[N-2])$ & \dots  & \dots          & \dots  &&&\\
\vdots          & \vdots          & \vdots & \vdots         & \vdots &&&\\
\dots           & \dots           & \dots  & $L(t,T_{k+1})$ &&&&\\
$L(t,T_3)$      & $L(t,\ts[3])$   & \dots  &&&&&\\
$L(t,T_2)$      &&&&&&&\\
  \end{tabular}~\\[1ex]
  \caption{Matrix of dependencies for LIBOR rates}
  \label{LIBOR-matrix}
 \end{center}
\end{table}}

The standard remedy to this problem is the so-called ```frozen drift''
approximation, where one replaces the random terms in the drift by
their deterministic initial values. This simplifies the simulations
considerably, since rates are no longer state-dependent and can be
simulated in parallel. However, this approximation is very crude
and does not yield acceptable results; cf. section \ref{numerics}.

Moreover, an additional numerical problem arises in LMMs with jumps
from the product term $\prod_l\beta(\cdot,\cdot,T_l)$. This term grows
exponentially as a function of tenor dates $N$ and makes the simulations
even more time-consuming.

\section{Picard approximation and drift expansion\\for the \lev \lib model}
\label{Picard-section}

The aim of this section is to derive approximation schemes for
\lib models that can overcome the pitfalls of the the model --
namely, the slow Monte Carlo simulations and the exponential
growth of the product term. Firstly, using the idea of Picard
iterations for the solution of SDEs, we derive approximate
equations for the dynamics of \lib rates which are suitable for
parallel computing. Secondly, by
expanding and truncating the product term we can reduce the
exponential to quadratic growth. Numerical examples show that
these methods yield significant gain in computational time, while
the loss in accuracy is very small.

\subsection{Picard iterations}

In order to derive approximation schemes for the dynamics of \lib
rates it is more convenient to work with the logarithm of rates.
Let us denote by $Z$ the log-\lib rates, that is
\begin{align}\label{log-LIB-SIE}
Z(t,T_i)
 &:= \log L(t,T_i) \nonumber\\
 &= Z(0,T_i) + \int_0^t b(s,T_i)\ud s + \int_0^t \volT\ud H_s,
\end{align}
where $Z(0,T_i)=\log L(0,T_i)$ for all $i\in\{1,\dots,N\}$. We can
immediately deduce that $Z(\cdot,T_i)$ is a semimartingale and its
triplet of predictable characteristics under $\pt[*]$,
$\mathbb{T}(Z(\cdot,T_i)|\P_{T_*})=(B^i,C^i,\nu^i)$, is described by
\begin{align}
 B^i &= \int\nolimits_0^\cdot b(s,\ts[i])\ds \nonumber\\
 C^i &= \int\nolimits_0^\cdot \vol2T c_s\ds\\
 1_A(x)*\nu^i &= 1_A\big(\volT x\big)*\nu^{T_*}, \qquad A\in\mathcal B(\Rmz).
\nonumber
\end{align}
The assertion follows from the canonical decomposition of a
semimartingale and the triplet of characteristics of the stochastic
integral process; see, for example, Proposition 1.3 in
\citeN{Papapantoleon06}.

\begin{remark}
Note that the martingale part of $Z(\cdot,T_i)$, i.e. the stochastic
integral $\int_0^\cdot\volT\ud H_s$, is a \tih \lev process. However,
the random drift term destroys the \lev property of $Z(\cdot,T_i)$,
as the increments are no longer independent.
\end{remark}

The dynamics of log-\lib rates can be alternatively described as
the solution to the following \emph{linear} SDE
\begin{align}\label{log-LIB-SDE}
\begin{split}
\ud Z(t,T_i) &= b(t,T_i;Z(t))\dt + \lambda(t,T_i)\ud H_t\\
    Z(0,T_i) &= \log L(0,T_i)
\end{split}
\end{align}
for all $i\in\{1,\dots,N\}$ and all $t\in[0,T_i]$. We have introduced
the term $Z(\cdot)$ in the drift term $b(\cdot,T_i;Z(\cdot))$ to make
explicit that the log-\lib rates depend on all subsequent rates in
the tenor.

The idea behind the Picard approximation scheme is to approximate the
dynamics of \lib rates by the Picard iterations for the SDE
\eqref{log-LIB-SDE}. The first Picard iteration for \eqref{log-LIB-SDE}
is simply the initial value, i.e.
\begin{align}\label{Picard-1}
Z^{(0)}(t,T_i) = Z(0,T_i),
\end{align}
while the second Picard iteration is
\begin{align}\label{Picard-2}
Z^{(1)}(t,T_i)
 &= Z(0,T_i) + \int_0^t b(s,T_i;Z^{(0)}(s))\ds + \int_0^t\lambda(s,T_i)\ud H_s
\nonumber\\
 &= Z(0,T_i) + \int_0^t b(s,T_i;Z(0))\ds + \int_0^t\lambda(s,T_i)\ud H_s.
\end{align}
Since the drift term $b(\cdot,T_i;Z(0))$ is deterministic, as the random terms
have been replaced with their initial values, we can easily deduce that the
second
Picard iterate $Z^{(1)}(\cdot,T_i)$ is a \lev process.

\begin{remark}
Comparing \eqref{Picard-2} with \eqref{log-LIB-SIE} it becomes evident
that we are approximating the semimartingale $Z(\cdot,T_i)$ with the
\tih \textit{\lev process} $Z^{(1)}(\cdot,\ts[i])$.
\end{remark}

\subsection{Application to LIBOR models}\label{Pic-LIBOR}

We will now use the Picard iterations in order to deduce strong -- i.e.
pathwise -- approximation schemes for the dynamics of \lib rates. More
specifically, we will use the Picard iterates as proxies for the
log-\lib rates in the drift term of the dynamics, cf. \eqref{LIBOR-drift-PT}.
Obviously, using the first Picard iterate $Z^{(0)}$ in the drift term
we have just recovered the ``frozen drift'' approximation.

Let us denote by $\widehat{Z}(\cdot,T_i)$ the approximate log-\lib rate
stemming from using the second Picard iterate $Z^{(1)}$. The dynamics
of the \textit{approximate} log-\lib rate is
\begin{align}\label{Picard-I}
\widehat{Z}(t,T_i)
 &= Z(0,T_i) + \int_0^t b(s,\ts[i];Z^{(1)}(s))\ds + \int_0^t \lambda(s,T_i)\ud
H_s,
\end{align}
where the drift term is provided by
\begin{align}\label{Picard-II}
b(s,\ts[i];Z^{(1)}(s))
 & = -\half \vol2T c_s
     - c_s\volT \sum_{l=i+1}^{N}
          \frac{\delta_l \e^{Z^{(1)}(s-,T_l)}}{1+\delta_l
\e^{Z^{(1)}(s-,T_l)}}\lambda(s,T_l)
     \nonumber\\&\,\,
     - \int_{\R}\left(\Big(\e^{\volT
x}-1\Big)\prod_{l=i+1}^{N}\widehat\beta(s,x,T_l)
                      -\volT x\right)F_s^{T_*}(\ud x),
\end{align}
with
\begin{align}\label{Picard-III}
\widehat\beta(t,x,T_l,)
 = \frac{\delta_l \exp\big(Z^{(1)}(t-,T_l)\big)}
        {1+\delta_l
\exp\big(Z^{(1)}(t-,T_l)\big)}\Big(\e^{\lambda(t,T_l)x}-1\Big) +1.
\end{align}

The main advantage of this Picard approximation is that the resulting SDE
for $\widehat{Z}(\cdot,T_i)$ can be simulated more easily than the equation
for $Z(\cdot,T_i)$. Indeed, contrary to the dynamics of $Z(\cdot,T_i)$,
the dynamics of $\widehat{Z}(\cdot,T_i)$ depend only on the \lev processes
$Z^{(1)}(\cdot,T_l)$, $i+1\le l\le N$, which are \emph{independent} of each
other. Compare also the ``dependence matrix'' for the approximate rates
(Table \ref{LIBOR-approx-matrix}) with Table \ref{LIBOR-matrix}.
Hence, we can use \textit{parallel computing} to simulate all
approximate \lib rates simultaneously. This significantly increases the
speed of the Monte Carlo simulations while, as the numerical examples reveal,
the empirical performance is very satisfactory.

{\footnotesize\renewcommand{\arraystretch}{1.35}
\begin{table}[h]
 \begin{center}
  \begin{tabular}{ccccccc}
\hline
$\widehat{L}(t,T_1)$  & $\widehat{L}(t,T_2)$  & \dots  & $\widehat{L}(t,T_k)$ &
\dots & $\widehat{L}(t,\ts[N-1])$ & $L(t,T_N)$\\
\hline
$Z^{(1)}(t,T_N)$      & $Z^{(1)}(t,T_N)$      & \dots  & $Z^{(1)}(t,T_N)$     &
\dots & $Z^{(1)}(t,T_N)$ &\\
$Z^{(1)}(t,\ts[N-1])$ & $Z^{(1)}(t,\ts[N-1])$ & \dots  & $Z^{(1)}(t,T_{N-1})$ &
\dots &&\\
$Z^{(1)}(t,\ts[N-2])$ & $Z^{(1)}(t,\ts[N-2])$ & \dots  & \dots                &
\dots &&\\
\vdots                & \vdots                & \vdots & \vdots               &
\vdots&&\\
\dots                 & \dots                 & \dots  & $Z^{(1)}(t,T_{k+1})$
&&&\\
$Z^{(1)}(t,T_3)$      & $Z^{(1)}(t,\ts[3])$   & \dots  &&&\\
$Z^{(1)}(t,T_2)$      &&&&&\\
  \end{tabular}~\\[1ex]
  \caption{Matrix of dependencies for approximate LIBOR rates}
  \label{LIBOR-approx-matrix}
 \end{center}
\end{table}}

\begin{remark}
Note that the Picard approximation can be also used in case one wants to apply
P(I)DE methods for the valuation of derivatives in \lib models, and yields an
analogous simplification of the problem.
\end{remark}

\begin{remark}
Let us point out that the Picard approximation scheme
\eqref{Picard-I}--\eqref{Picard-III} we have developed is \emph{universal} and
can be applied to any LMM. We can replace the \lev process $H$ driving the
dynamics of \lib rates by a general semimartingale $X$ with random
predictable characteristics (thus also incorporating stochastic
volatility). Subject to certain assumptions $X$ has the canonical
decomposition
\begin{align}
X_t = \int_0^t \sqrt{c_s}\ud W_s + \int_0^t\int_{\R}x(\mu^H-\nu)\dsdx;
\end{align}
compare with \eqref{LIBOR-Levy}. Then we can construct an LMM driven by the
semimartingale $X$ following the steps for the \lev LMM in Eberlein and \"Ozkan
\citeyear{EberleinOezkan05}. We can also follow analogously all the steps for the
Picard approximation; in this case, the Picard iterate $Z^{(1)}$ will have the
same dynamics as in \eqref{Picard-2}, with $H$ replaced by $X$.
\end{remark}

\subsection{Drift expansion}
\label{driftsecond}

This part is devoted to the integral term in the drift of the \lib
dynamics; see \eqref{LIBOR-drift-PT} again. Obviously this is a
problem solely related to LMMs driven by jump processes.

Let us introduce the following shorthand notation for convenience:
\begin{align}
 \lambda_l := \lambda(s,T_l)
 \quad\text{ and }\quad
 L_l := L(s,T_l).
\end{align}
We denote by $\mathbb{A}$ the part of the drift term that stems from the
jumps, i.e.
\begin{align}
\mathbb{A}
 &= \int_{\R}\left( \Big(\e^{\lambda_i x}-1\Big)
      \prod_{l=i+1}^{N} \left(1+\frac{\delta_l L_l}{1+\delta_l L_l}
            \Big(\e^{\lambda_lx}-1\Big)\right)-\lambda_i x\right)F_s^{T_*}(\ud
x).
\end{align}
In theory, one could simply employ a straightforward numerical
integration to compute $\mathbb{A}$. However this is not feasible
in practice since a numerical integration should be performed at
each step of the Monte Carlo simulation.
An alternative solution is to express $\mathbb{A}$ in terms of the
cumulant generating function of (the jump part) of $H$.

Observe that a product of the form $\prod_{l=1}^N(1+\alpha_l)$ appears, where
$\alpha_l:=\frac{\delta_l L_l}{1+\delta_l L_l}(\e^{\lambda_lx}-1)$. This
product can be expressed in terms of so-called \emph{elementary symmetric
polynomials}. Let $k\le N$, then the elementary symmetric polynomial
of degree $k$ in $N$ variables is given by
\begin{align}\label{ell-sym-pol}
\varepsilon_k(\alpha_1,\dots,\alpha_N)
 = \sum_{1\le i_1<\dots<i_k\le N} \alpha_{i_1}\times\dots\times \alpha_{i_k}.
\end{align}
Hence we have that
\begin{align}\label{product}
\prod_{l=1}^N(1+\alpha_l)
 = 1 + \varepsilon_1(\alpha_1,\dots,\alpha_N) + \dots +
\varepsilon_N(\alpha_1,\dots,\alpha_N).
\end{align}

One can immediately deduce that, while the drift term stemming from the
diffusion part is a first order polynomial in $\frac{\delta_l L_l}{1+\delta_l
L_l}$,
$\mathbb{A}$ is an $N$-th order polynomial. More importantly, the number
of terms on the RHS of \eqref{product} is $2^N$. Hence, we need to
perform $2^N$ computations in order to calculate the drift of the \lib rates.
Since $N$ is the length of the tenor, it becomes apparent that this calculation
is feasible only for short tenors. If, for example, $N=40$ this
amounts to more than 1 trillion computations.

In order to make this computation more feasible we will truncate the RHS
of \eqref{product} at the first or second order. The first order
approximation of the product term is
\begin{align}
\label{drifta}
\mathbb{A}'
 &= \int_{\R} \left( \Big(\e^{\lambda_i x}-1\Big)
       \Big( 1 + \varepsilon_1(\alpha_{i+1},\dots,\alpha_N)-\lambda_i x\Big)
\right)F_s^{T_*}(\ud x) \nonumber\\
 &= \int_{\R} \left( \Big(\e^{\lambda_i x}-1\Big)
       \left( 1 + \sum_{l=i+1}^{N}\frac{\delta_l L_l}{1+\delta_l L_l}
            \Big(\e^{\lambda_lx}-1\Big)\right)-\lambda_i x\right)F_s^{T_*}(\ud
x) \nonumber\\
 &= \int_{\R} \left(\e^{\lambda_i x}-1-\lambda_i x\right)F_s^{T_*}(\ud x)
\nonumber\\
 &\quad + \sum_{l=i+1}^{N}\frac{\delta_l L_l}{1+\delta_l L_l}
    \int_{\R} \Big(\e^{\lambda_i
x}-1\Big)\Big(\e^{\lambda_lx}-1\Big)F_s^{T_*}(\ud x)\nonumber\\
 &= \kappa\big(\lambda_i\big)
  + \sum_{l=i+1}^{N}\frac{\delta_l L_l}{1+\delta_l L_l}
    \Big(\kappa\big(\lambda_i+\lambda_l\big)
           - \kappa\big(\lambda_i\big) - \kappa\big(\lambda_l\big)\Big),
\end{align}
and the order of the error is
\begin{align}
\mathbb{A} = \mathbb{A}' + O\big(N\|L\|^2\big).
\end{align}
Similarly the second order approximation is provided by
\begin{align}
\label{drifta2}
\mathbb{A}''
 &= \int_{\R} \left( \Big(\e^{\lambda_i x}-1\Big)
       \Big( 1 + (\varepsilon_1 + \varepsilon_2)(\alpha_{i+1},\dots,\alpha_N)
-\lambda_i x\Big)
                 \right)F_s^{T_*}(\ud x) \nonumber\\
 &= \kappa\big(\lambda_i\big)
  + \sum_{l=i+1}^{N}\frac{\delta_l L_l}{1+\delta_l L_l}
    \Big(\kappa\big(\lambda_i+\lambda_l\big)
           + \kappa\big(\lambda_i\big) + \kappa\big(\lambda_l\big)\Big)
\nonumber\\
 &\quad + \sum_{i+1\le k<l\le N}
           \frac{\delta_l L_l}{1+\delta_l L_l}\frac{\delta_k L_k}{1+\delta_k
L_k} \nonumber\\
 &\quad\qquad\times \Big( \kappa\big(\lambda_i+\lambda_l+\lambda_k\big)
        - \kappa\big(\lambda_i+\lambda_l\big)
        - \kappa\big(\lambda_i+\lambda_k\big) \nonumber\\
 &\qquad\quad\qquad        - \kappa\big(\lambda_k+\lambda_l\big)
        + \kappa\big(\lambda_i\big)
        + \kappa\big(\lambda_l\big)
        + \kappa\big(\lambda_k\big) \Big),
\end{align}
and the order of the error is
\begin{align}
\mathbb{A} = \mathbb{A}'' + O\big(N^2\|L\|^3\big).
\end{align}
Since the \lib rate is an order of magnitude smaller than the
number of tenor dates, these approximations are justified.
Indeed, numerical results show that truncation at the second order yields
very satisfying results, while the gain in computational time is
very significant.

\section{Derivatives on \lib rates}
\label{derivatives}

In this section we briefly describe the derivatives we will use
for the numerical illustration. Namely we use caplets and swaptions,
which are the most liquid derivatives in the interest rate markets.
We will also use forward rate agreements (FRAs) as a benchmark for
the different approximations because they have model independent
values. Of course, since we have developed a pathwise approximation,
we could also consider many other derivatives -- especially
path-dependent ones -- for the illustration. We avoid this for the sake
of brevity.

\subsection{Caplets}

The price of a caplet with strike $K$ maturing at time $T_i$, using
the relationship between the terminal and the forward measures
\eqref{TF-libor}, can be expressed as
\begin{align}\label{caplet}
\mathbb{C}_0(K,\ts[i])
 &= \delta_i B(0,\ts[i+1])\, \E_{\pt[i+1]}[(L(T_i,T_i)-K)^+] \nonumber\\
 &= \delta B(0,\ts[i+1])\,
\E_{\pt[*]}\Big[\frac{\ud\MeaT}{\ud\P_{T_*}}\big|_{\F_{\ts[i]}}(L(T_i,
T_i)-K)^+\Big] \nonumber\\
 &= \delta B(0,\ts[*])\,
    \E_{\pt[*]}\Big[\prod_{l=i+1}^{N}\big(1+\delta
L(T_i,T_l)\big)(L(T_i,T_i)-K)^+\Big].
\end{align}
This equation will provide the actual prices of caplets corresponding
to simulating the full SDE (Euler discretization) for the LIBOR rates.
In order to calculate the Picard approximation prices for a caplet we
have to replace $L(\cdot,T_\cdot)$ in \eqref{caplet} with
$\widehat{L}(\cdot,\ts[\cdot])$.

\subsection{Swaptions}

A payer (resp. receiver) swaption can be viewed as a put (resp. call)
option on a coupon bond with exercise price 1; cf. section 16.2.3 and
16.3.2 in \citeN{MusielaRutkowski97}. Consider a payer swaption with
strike rate $K$, where the underlying swap starts at time $T_i$ and
matures at $T_m$ ($i<m\le N$). The time-$T_i$ value is
\begin{align}
\mathbb{S}_{T_i}(K,T_i,T_m)
 &= \left( 1-\sum^m_{k=i+1} c_k B(T_i,T_k)\right)^+ \nonumber\\
 &= \left( 1-\sum^m_{k=i+1} \bigg(c_k \prod_{l=i}^{k-1}\frac{1}{1+\delta
L(T_i,T_l)}\bigg)\right)^+,
\end{align}
where
\begin{align}
c_k = \left\{
        \begin{array}{ll}
          \delta_k K, & \hbox{$i+1\le k\le m-1$,} \\
          1+\delta_k K, & \hbox{$k=m$.}
        \end{array}
      \right.
\end{align}
Then, the time-0 value of the swaption is obtained by taking the
$\pt[i]$-expectation of its time-$T_i$ value, that is
\begin{align}
\mathbb{S}_{0}
 &= \mathbb{S}_{0}(K,T_i,T_m) \nonumber\\
 &= B(0,T_i)\,
    \E_{\pt[i]}\left[\left( 1-\sum^m_{k=i+1}
       \bigg(c_k \prod_{l=i}^{k-1}\frac{1}{1+\delta
L(T_i,T_l)}\bigg)\right)^+\right] \nonumber\\
 &= B(0,T_*)\,\nonumber\\
 &\quad\times    \E_{\pt[*]}\left[\prod_{l=i}^{N}\big(1+\delta L(T_i,T_l)\big)
               \left( 1-\sum^m_{k=i+1}
               \bigg(c_k \prod_{l=i}^{k-1}\frac{1}{1+\delta
L(T_i,T_l)}\bigg)\right)^+\right], \nonumber
\end{align}
hence
\begin{align}\label{swap-1}
\mathbb{S}_{0} &= B(0,T_*)\,
    \E_{\pt[*]}\left[\left( - \sum^m_{k=i}
               \bigg(c_k \prod_{l=k}^{N}\left(1+\delta
L(T_i,T_l)\right)\bigg)\right)^+\right],
\end{align}
where $c_i:=-1$.

\subsection{Forward Rate Agreements}

A forward rate agreement with strike $K$ and notional value of 1
with expiry at time $T_i$, has the value of $\delta_i(K-L(T_i,T_i))$
at expiry. The time zero value of the contract has the model
independent value of
\begin{align}
\mathbb{F}_0(K,\ts[i])
 = \delta_i B(0,\ts[i+1])[K-L(0,T_i)],
\end{align}
or zero if the contract is struck at-the-money $(K=L(0,T_i))$. In
the following section we will compare the known true values with
simulated prices generated from the terminal measure expectation
of the payoff, i.e.
\begin{align}\label{FRA}
\mathbb{F}_0(K,\ts[i])
 &= \delta B(0,\ts[*])\,
    \E_{\pt[*]}\Big[\prod_{l=i+1}^{N}\big(1+\delta
L(T_i,T_l)\big)(K-L(T_i,T_i))\Big].
\end{align}

\section{Numerical illustration}
\label{numerics}

The aim of this section is to demonstrate the accuracy and efficiency of the
Picard approximation scheme and the drift expansions in the valuation of options
in the \lev \lib model. In the first section we demonstrate the accuracy of our
methods compared to a standard Euler discretization of the LIBOR SDE in pricing
caplets and swaptions. In the second section we estimate the speed of our
method; finally we compare with alternative approximations.

We will consider a simple example with a flat volatility structure of
$\lambda(\cdot,T_i)=18\%$ and zero coupon rates generated from a flat term
structure of interest rates: $B(0,T_i)=\exp(-0.04\cdot T_i)$. We consider a
tenor structure with 6 month increments ($\delta_i=\frac12$).

The driving \lev process $H$ is a normal inverse Gaussian (NIG) process with
parameters $\alpha=\bar\delta=12$ and $\mu=\beta=0$, resulting in a process with
mean zero and variance 1. We denote by $\mu^H$ the random measure of jumps of
$H$ and by $\nu\dtdx=F(\dx)\dt$ the $\P_{T_*}$-compensator of $\mu^H$, where $F$
is the \lev measure of the NIG process. The necessary conditions are then
satisfied for term structures up to 30 years of length because $M=\alpha$, hence
$\sum_{i=1}^{60}|\lambda(\cdot,T_i)|=10.8<\alpha$.

The NIG \lev process is a pure-jump process with canonical decomposition
\begin{align}
H=\int_0^\cdot\int_\R x (\mu^H-\nu)\dsdx.
\end{align}
The cumulant generating function of the NIG distribution, for all
$u\in\mathbb{C}$ with $|\Re u|\le\alpha$, is
\begin{align}
\kappa(u)
 &= \bar\delta\alpha-\bar\delta\sqrt{\alpha^2-u^2}.
\end{align}

\subsection{Accuracy of the methods}

The Picard approximation should be considered primarily as a parallelizable
alternative to the standard Euler discretization of the model. This will
therefore be the benchmark to which we compare. In order to avoid Monte Carlo
error we use the same discretization grid (5 steps per tenor increment) and the
same pseudo random numbers (50000 paths) for each method. The pseudo random
numbers are generated from the NIG distribution using the standard methodology
described in \citeN{Glasserman03}.

\begin{figure}[ht!]
 \centering
 \includegraphics[width=6.25cm]{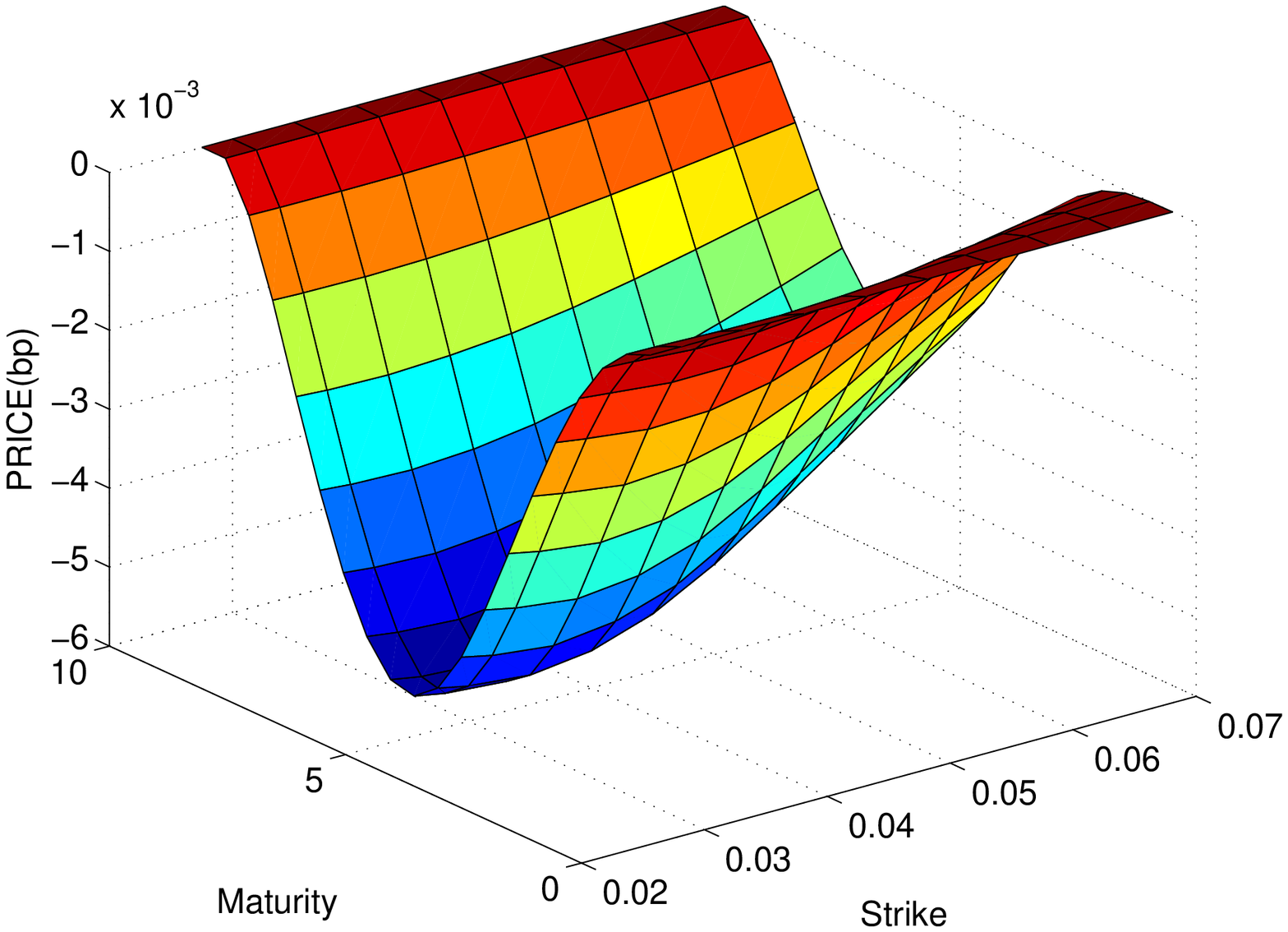}
 \includegraphics[width=6.25cm]{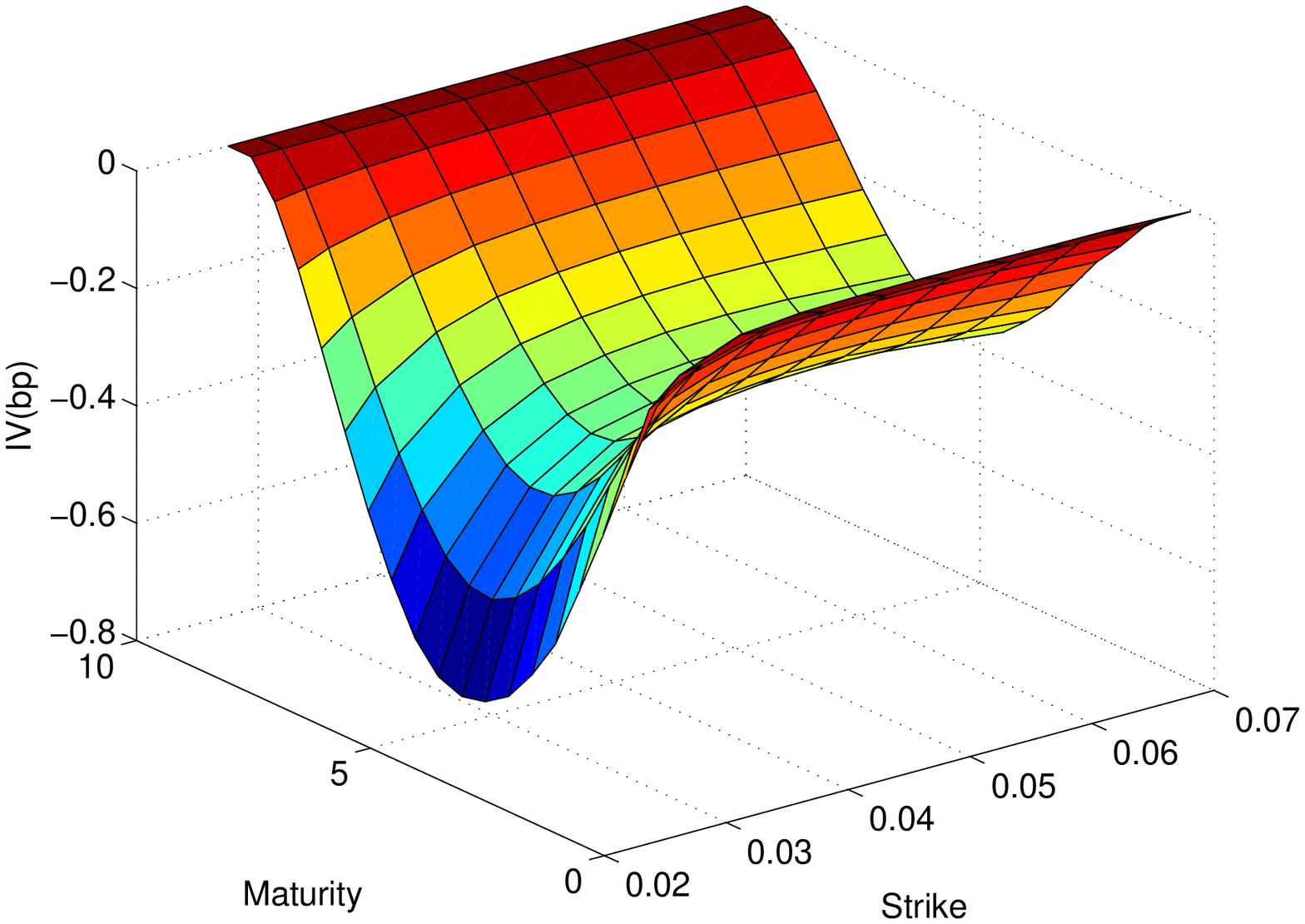}
 \caption{Difference in price (left) and implied caplet volatility (right)
          between the Euler discretization and the Picard approximation (in
          basis points).}
 \label{fig:caplets-diffs}
\end{figure}

Starting with caplets we can see in Figure \ref{fig:caplets-diffs} the
difference between the Euler discretization and the Picard approximation
expressed in price (left) and implied volatility (right). The difference in
price is small with max. errors around half of a percentage of a basis point. On
the right we see somewhat larger errors with a maximum slightly below 1 basis
point of implied volatility for low strike mid-maturity caplets. Implied
volatility is normally quoted in units of 1 basis point while bid-ask spreads
are usually around at least 5 bp of implied volatility. The errors are therefore
at acceptable levels. Note also that in experiments not shown we found that the
levels and patterns of the errors are insensitive to the number of
discretization points as well the number of paths.

The errors display a non-monotonic behavior as a function of maturity with peaks
around mid-maturity. The non-monotonicity can be explained by the fact that
volatility dominates the price of options in the short end, making the drift,
and any error in it, less relevant. As maturity increases the importance of the
drift grows relative to volatility but the state dependence becomes less
critical as the number of ``live" rates decreases. These two opposing effects
result in the mid-maturity peak that we observe. This pattern is also noted in
the study by \citeN{JoshiStacey08}.

\begin{figure}[ht!]
 \centering
 \includegraphics[width=7cm]{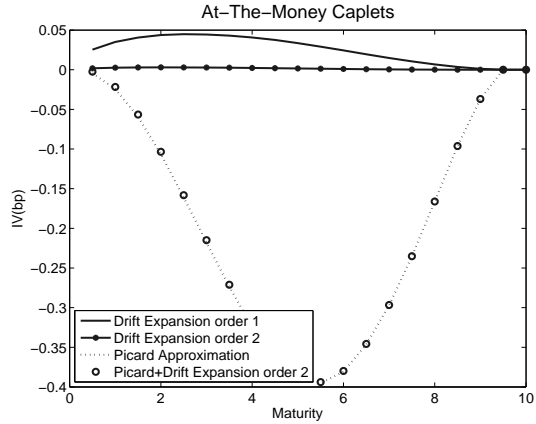}
 \caption{Difference in implied caplet volatility (in basis points) between
          Euler discretization and 4 other methods.}
 \label{fig:ATMcaplets-diffs}
\end{figure}

As we established in \eqref{product} the number of terms needed to calculate the
drift grows at a rate $2^N$. In market applications $N$ is often as high as 60
reflecting a 30 year term structure with a 6 month tenor increment. At this
level even the calculation of one drift term becomes infeasible and this
necessitates the use of the approximations introduced in \eqref{drifta} and
\eqref{drifta2}. We investigate the errors introduced by the drift expansion by
comparing them with the full numerical solution obtained as before using the
true drift from  \eqref{LIBOR-drift-PT}. The results are plotted in Figure
\ref{fig:ATMcaplets-diffs}. Here we can see that the first order drift expansion
adds errors of fairly low magnitude, whereas the second order drift expansion
performs significantly better with errors so small that they can be disregarded.
We also plot the Picard approximation alone as well as combined with the second
order drift expansion and these are similarly indistinguishable.

\begin{figure}[ht!]
 \centering
 \includegraphics[width=6.25cm]{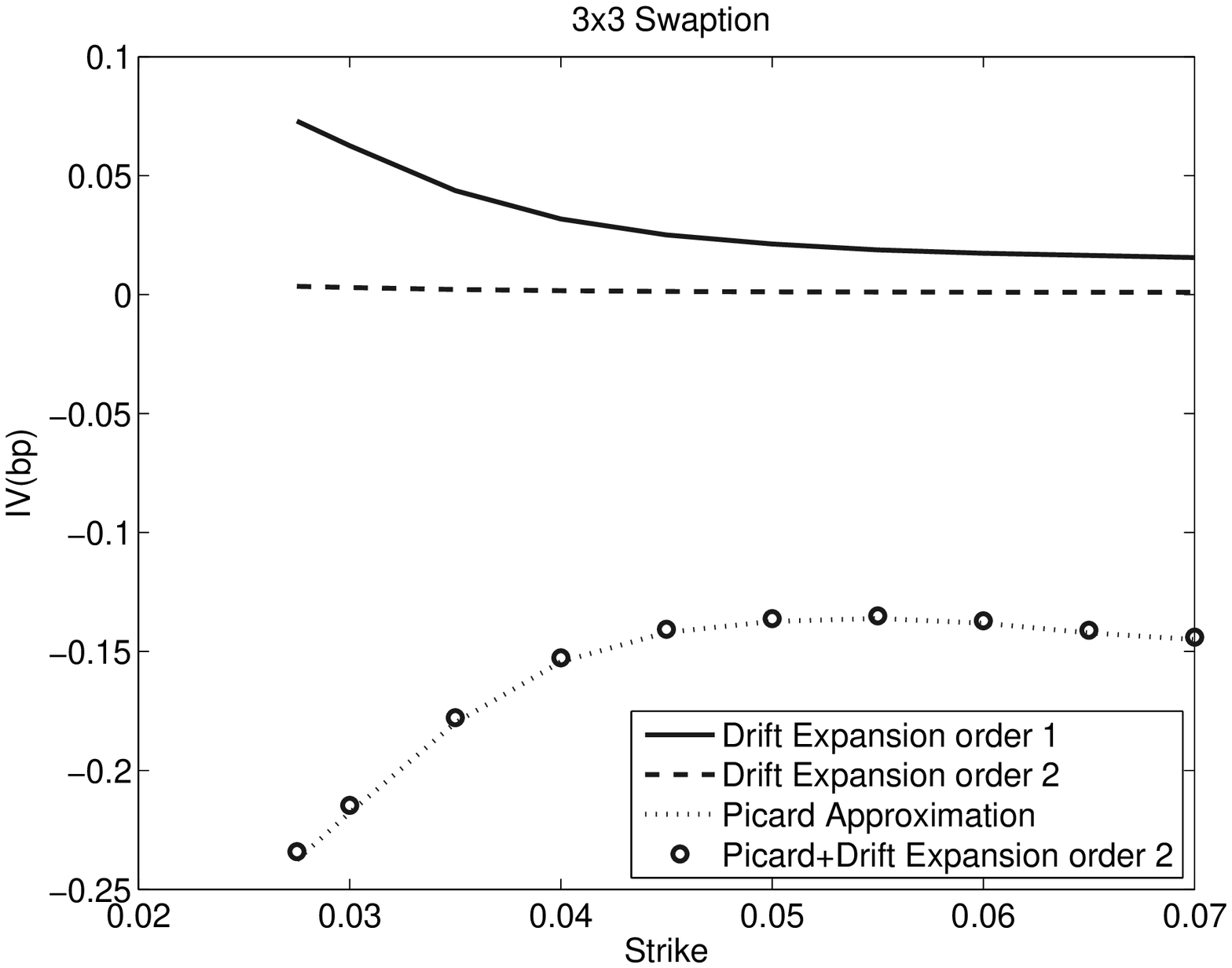}
 \includegraphics[width=6.25cm]{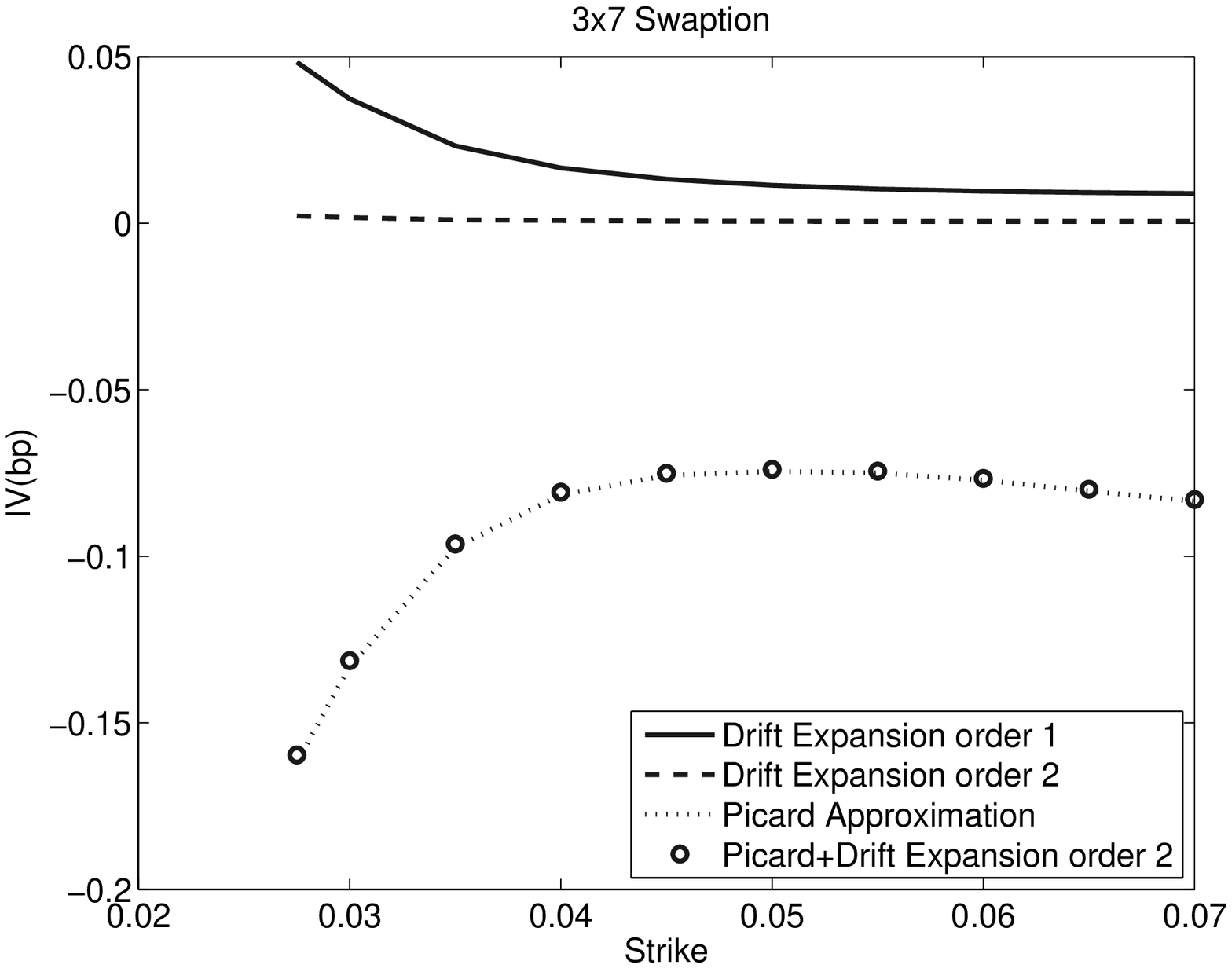}
 \includegraphics[width=6.25cm]{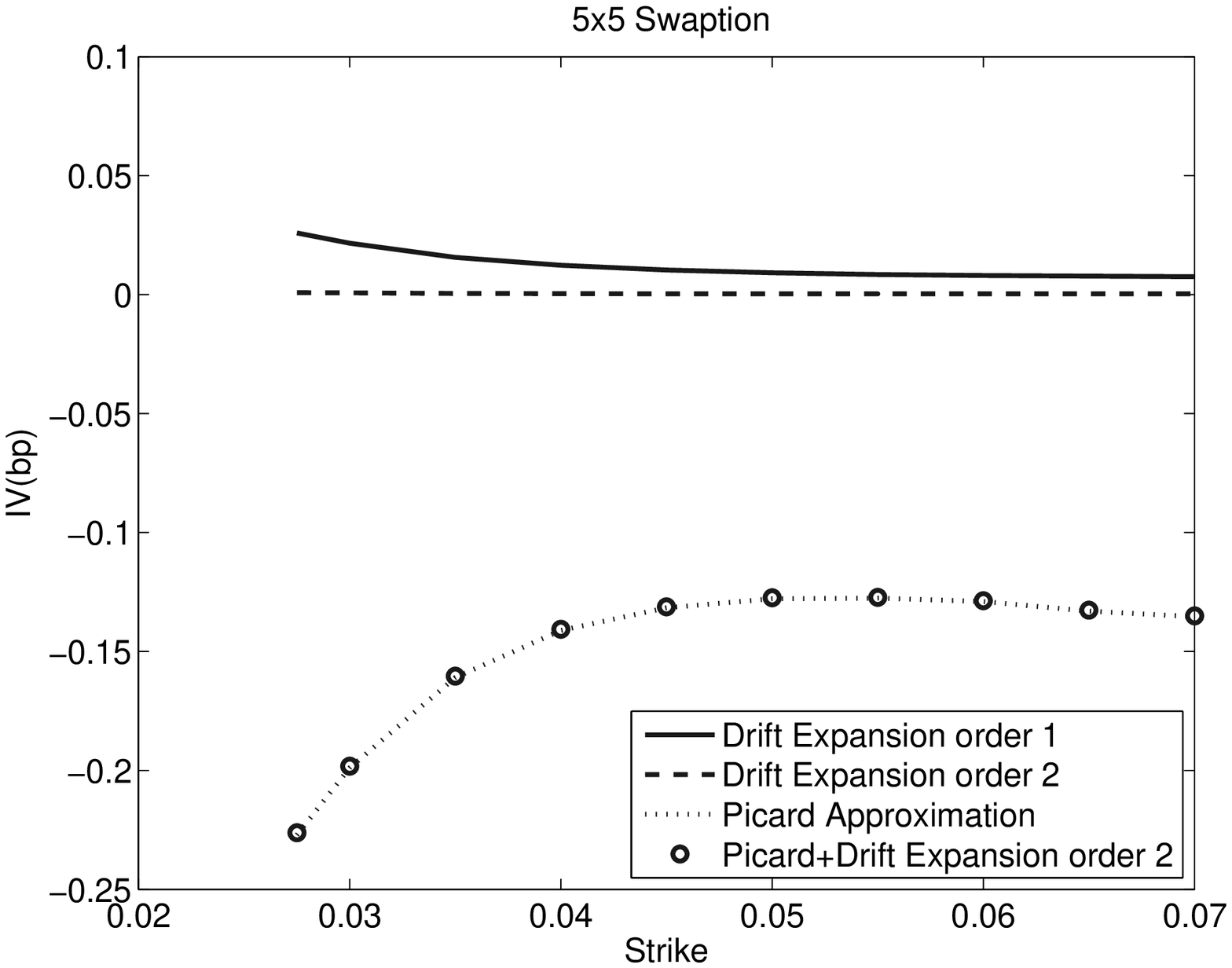}
 \caption{Difference in implied swaption volatility (basis points) between Euler
          discretization and 4 other methods.}
 \label{fig:swaptions-diffs}
\end{figure}

Continuing on with swaptions, in Figure \ref{fig:swaptions-diffs} we plot again
differences in implied volatility between our methods and the Euler
discretization. We observe even smaller errors than for caplets, most likely
because swaptions span a broad range of maturities which smooths out the higher
errors in the mid-maturity region.


\begin{figure}[ht!]
 \centering
 \includegraphics[width=6.25cm]{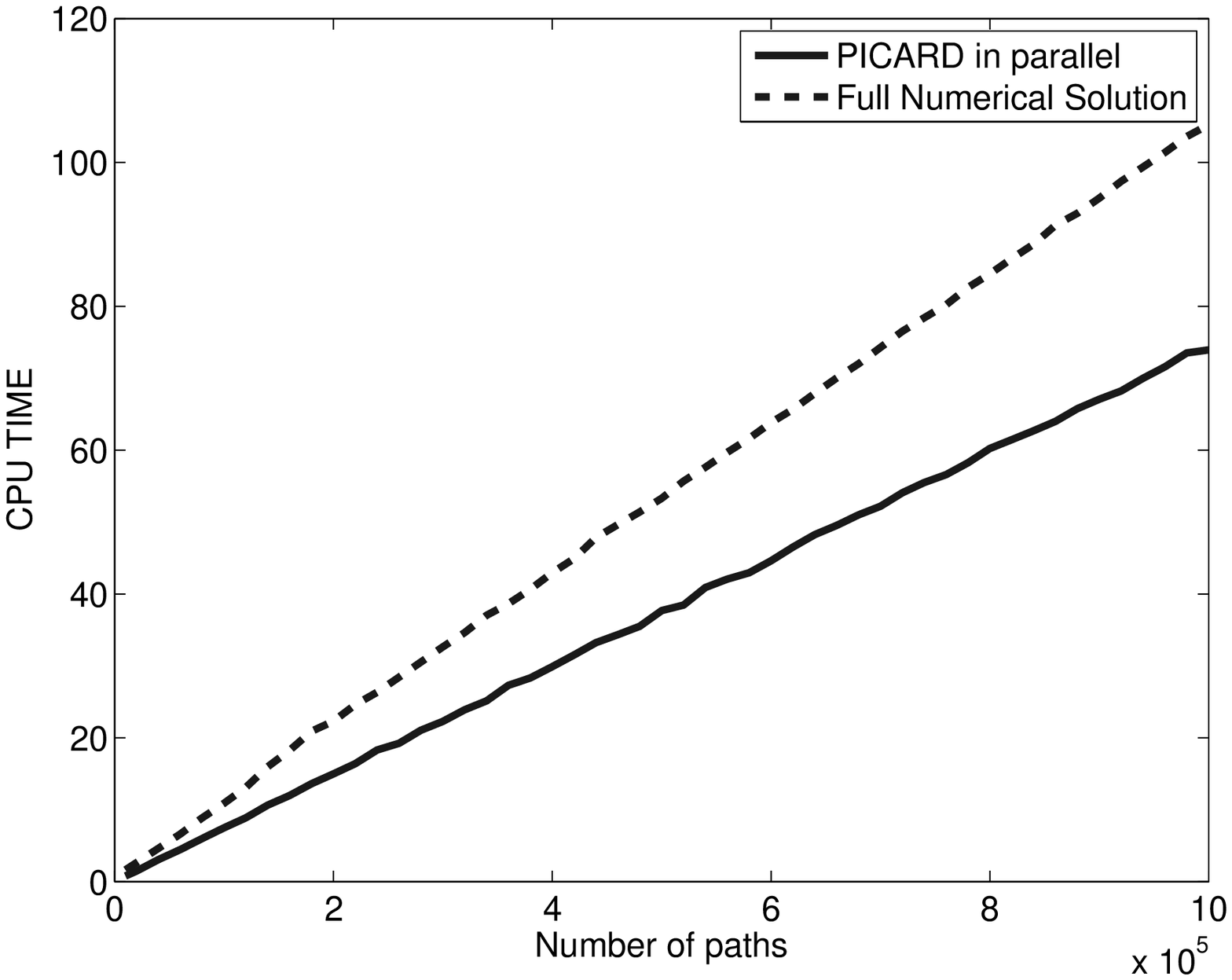}
 \includegraphics[width=6.25cm]{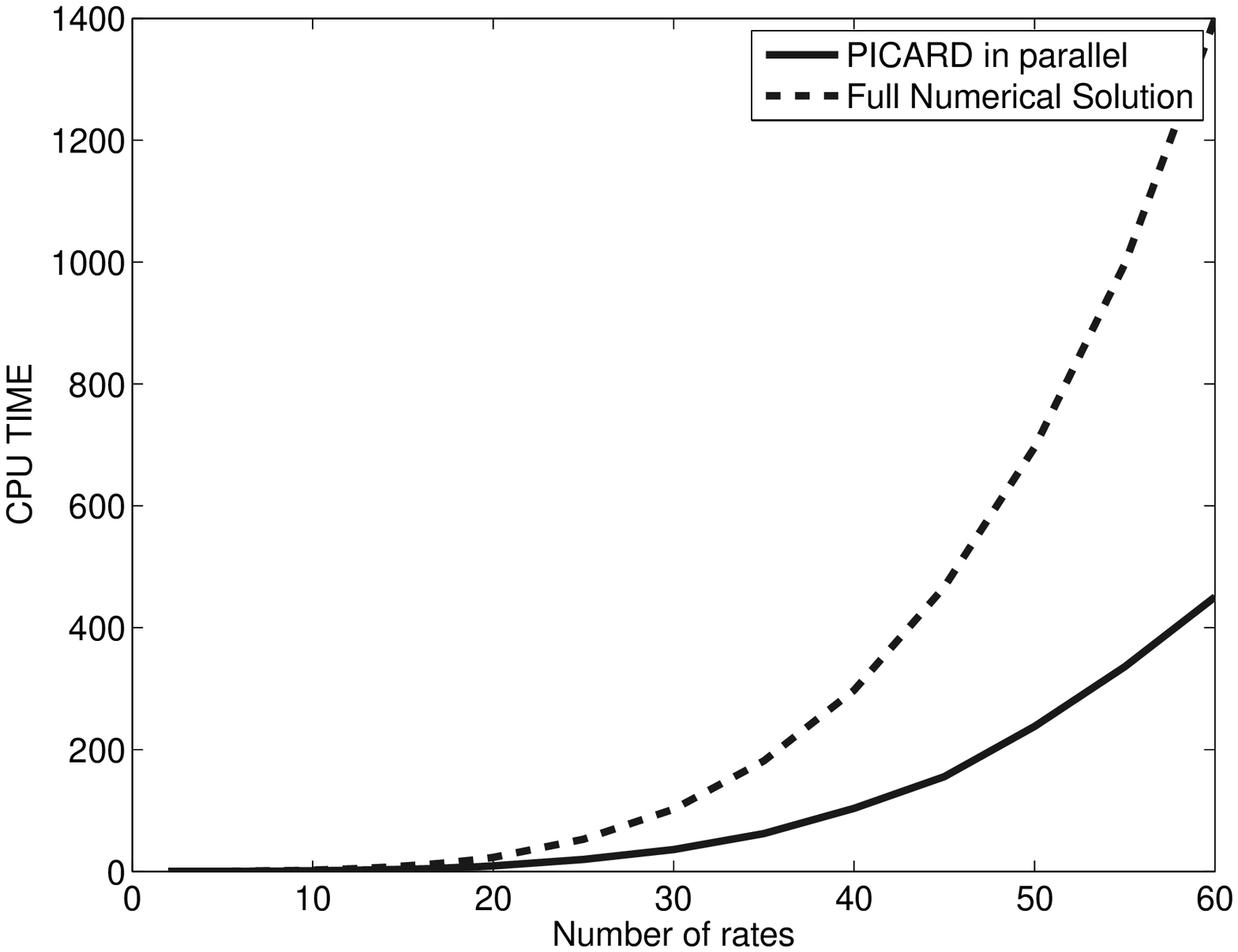}
 \caption{CPU time as a function of the number of paths (left) and the number of
          rates $N$ (right).}
 \label{fig:caplets-CPU-TIME}
\end{figure}

\subsection{Computational speed}

In terms of computational time the largest gain by far is realized when using
the drift expansions in \eqref{drifta} and \eqref{drifta2}. In the example above
the CPU time for the full numerical solution is 1.5 hours; after applying the
first order and second order drift expansion it drops to 1.3 and 27.2 seconds
respectively.

The Picard approximation in itself does not contribute to the computational
speed unless parallelization is employed. In fact, it is slightly but
insignificantly slower as the auxiliary processes $Z^{(1)}$ have to be evolved
along with the rates. On the left in Figure \ref{fig:caplets-CPU-TIME}, we plot
CPU time as a function of the number of paths for the Picard approximation and
the full Euler discretization. In both cases the second order drift
approximation scheme in \eqref{drifta2} is employed. The computations are done
in Matlab running on an Intel i7 processor with the capability of running 8
processes simultaneously. Here we see the typical linear behavior as the number
of paths are increased; notice though that the Picard approximation has
significantly lower slope. On the right in Figure \ref{fig:caplets-CPU-TIME} we
plot CPU time as a function of the number of rates. One can see CPU time
exponentially increasing, revealing that large gains in computational time are
realizable when using the Picard approximation scheme and the drift expansion.

Needless to say, the speed advantages of the Picard approximation are only
partially revealed in these graphs since we are using desktop computer; further
speed increases can of course be realized as the access to more CPUs (or
clusters of PCs) becomes available. This is already part of the infrastructure
of many large financial institutions.

\subsection{Comparison with other methods}

Unfortunately, very little work has been done in the area of approximations for
LMMs driven by general semimartingales, leaving us without any standard method
to compare with, other than the frozen drift approximation. As mentioned in the
introduction the existing work has focused mainly on the log-normal case and the
case of finite intensity jump-diffusion models. However, some of the techniques
applied to the log-normal case can be adapted to our setup as well.

Assume we want to simulate the \lib rates from time $t$ to time $t+h$,
where $t+h\leq T_i$. We have
\begin{align*}
L(t+h,T_i) = L(t,T_i)
 \exp\left( \int_t^{t+h} b(s,T_i;Z(s))\ud s+\int_t^{t+h} \volT\ud H_s
\right),
\end{align*}
where $b(\cdot,T_i;Z(\cdot))$ is the state dependent drift function defined in
\eqref{LIBOR-drift-PT}; cf. also \eqref{log-LIB-SDE}. The standard (log)-Euler
scheme leads to
\begin{align}
\int_t^{t+h} b(s,T_i;Z(s))\ud s \approx  b(t,T_i;Z(t)) \, h.
\end{align}
This can be further refined as noted first in \citeN{KloedenPlaten99} (see also
\citeNP{HunterJaeckelJoshi01}, and \citeNP{JoshiStacey08} and the references
therein) by using instead
\begin{align}\label{PC}
\int_t^{t+h} b(s,T_i;Z(s))\ud s \approx
\left(\frac{1}{2}b(t,T_i;Z(t))+\frac{1}{2}b(t+h,T_i;Z(t+h))\right)h.
\end{align}
The second term in the parenthesis requires the knowledge of the LIBOR rates
$L(t+h,T_{i+1}),\dots,L(t+h,T_N)$; one therefore has to simulate
these rates. This can be done in a separate simulation step and the procedure is
known in the LMM literature as \emph{predictor-corrector} (PC) method. One can
also note that when rate $i$ is evolved under the terminal measure it only
depends on rates $k>i$. Furthermore, if we start with $i=N$ we have no
state-dependence in the drift. We can then generate realizations of $L(t+h,T_N)$
without discretization error and these can be used in the drift of rate
$N-1$ as described above. Realizations of rate $N-1$ can then be generated with
the corrected drift from \eqref{LIBOR-drift-PT}, which can be subsequently used
in the drift of rate $N-2$, and so forth. This latter method is referred to as
\emph{iterative predictor-corrector} (IPC).

IPC has been found to often outperform PC in the log-normal case studied in
\citeN{JoshiStacey08}. It is also slightly more efficient since it does not
require a separate simulation step for the rates at time $t+h$.

\begin{remark}
One should point out that PC and IPC are alternative \textit{discretization
schemes} to the Euler discretization.
\end{remark}

We can also combine PC and IPC with the Picard approximation by merely using
$b(\cdot,T_i;Z^{(1)}(\cdot))$ instead of $b(\cdot,T_i;Z(\cdot))$ in \eqref{PC}.
Furthermore, PC and IPC will actually be equivalent when applied to the Picard
approximation since the drift term does not involve the rates themselves, but
the auxiliary processes $Z^{(1)}$, which makes the order in which the rates are
evolved irrelevant (see section \ref{Pic-LIBOR}).

\begin{figure}[ht!]
 \centering
 \includegraphics[width=6.25cm]{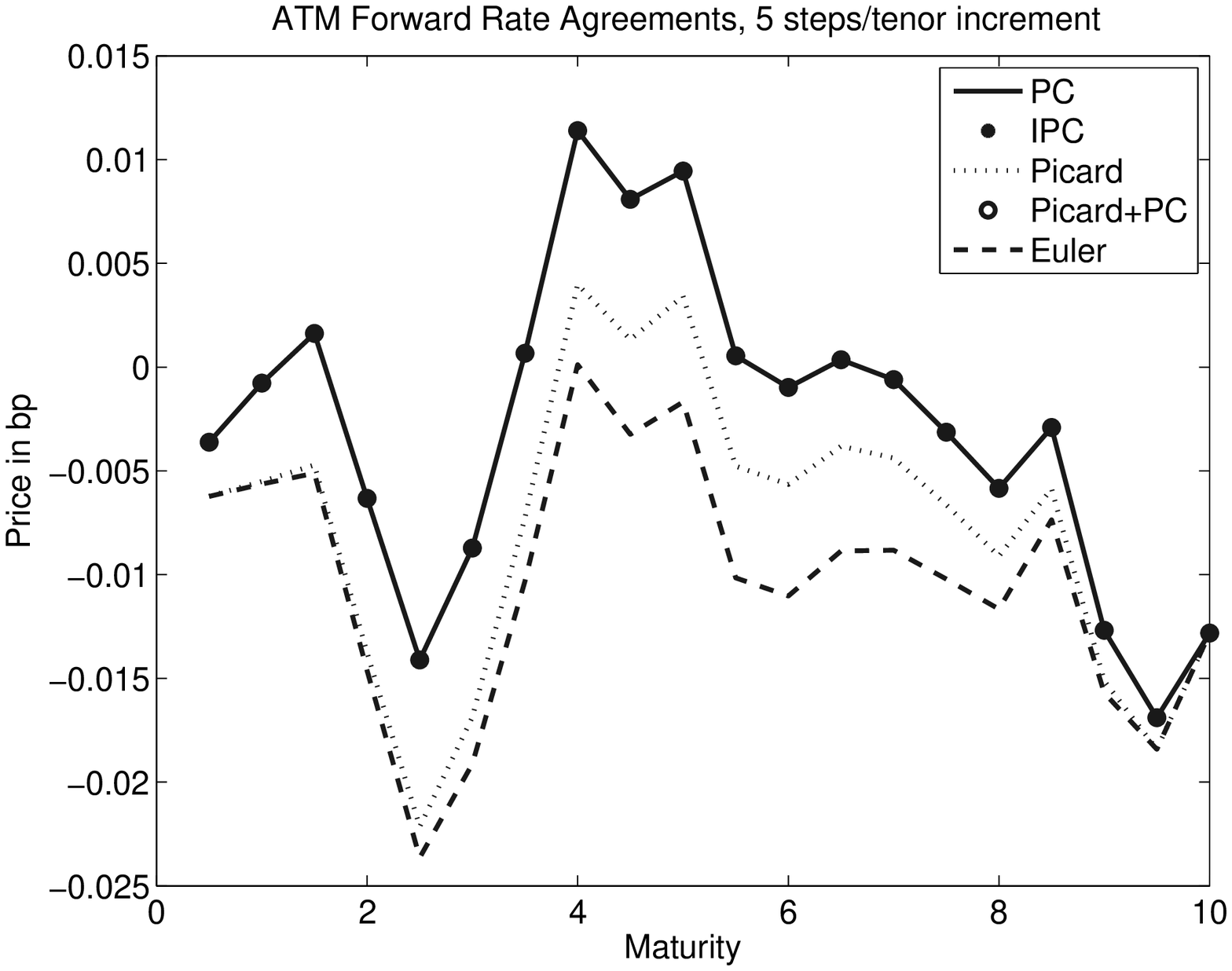}
 \includegraphics[width=6.25cm]{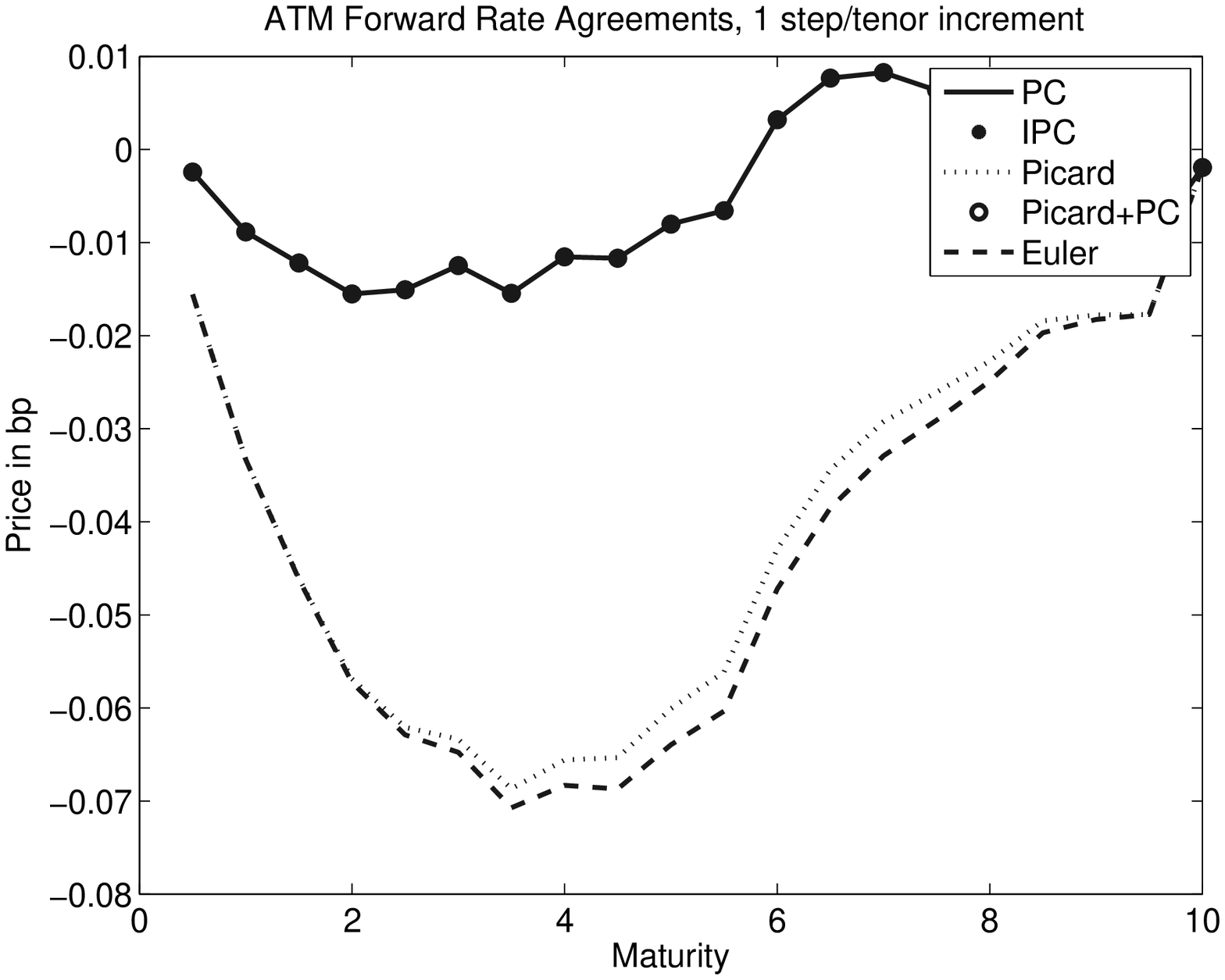}
 \includegraphics[width=6.25cm]{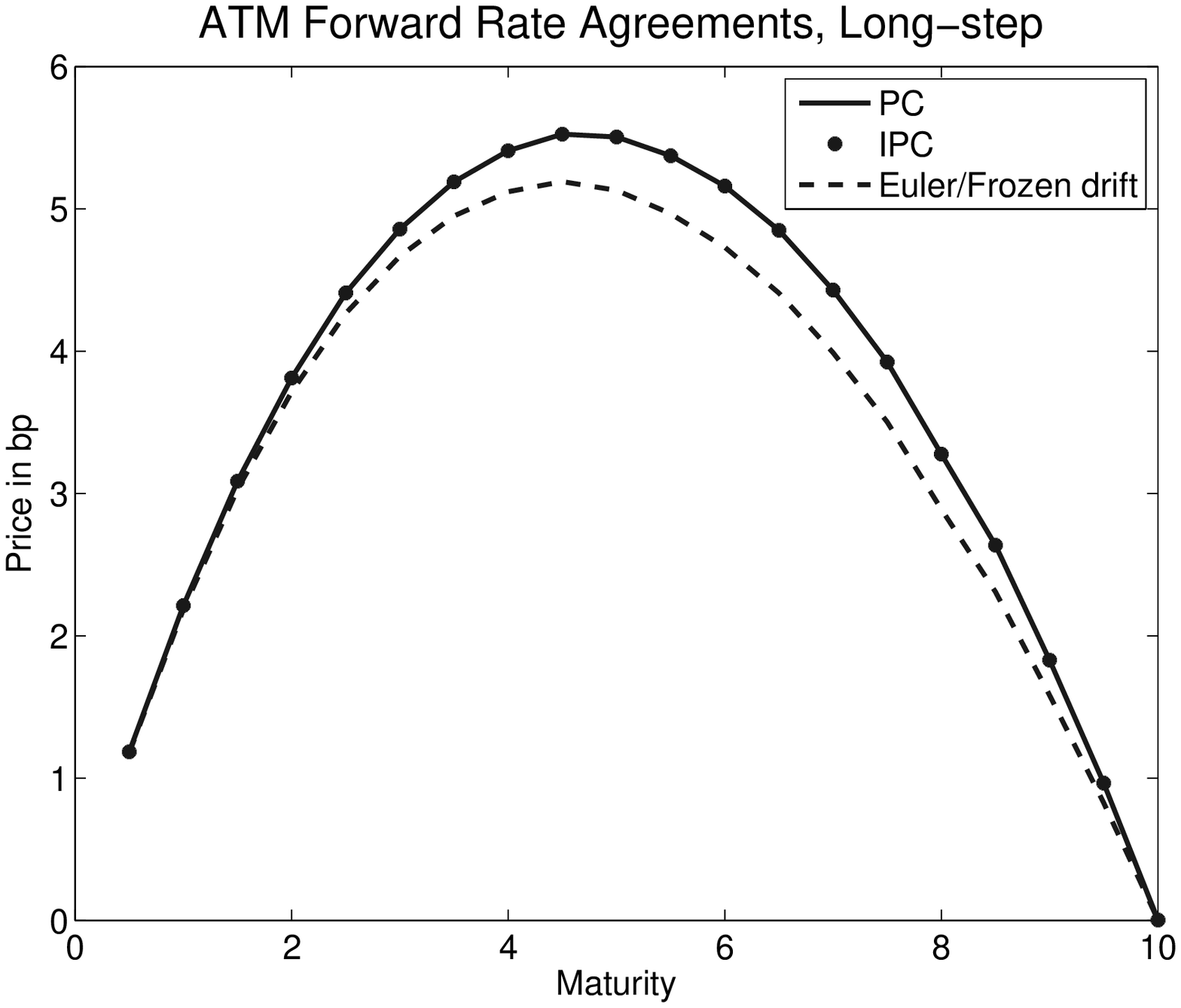}
  \caption{Prices for at-the-money Forward Rate Agreements (in basis points).
           The true price is zero.}
 \label{fig:PFRA}
\end{figure}

We compare PC and IPC with our methods in 3 different cases. Since we do not
have the true price of caplets or swaptions we instead compare prices of
at-the-money forward rate agreements (FRA) since these all have a known model
independent price of zero. Here we keep Monte Carlo error sufficiently low by
simulating 5 million paths, employing antithetic sampling as the only variance
redaction measure. Looking at the top left of Figure \ref{fig:PFRA} we have
kept the discretization grid dense at 5 steps per tenor period.  Here we see
prices which for all methods are sufficiently small at max levels of 2.5\% of
a basis point. Furthermore the different methods are more or less
indistinguishable -- certainly in statistical terms with 95\% confidence limit
halfwidths ranging from about 0.004 bp in the short end to to 0.03 basis points
in the long end. The frozen drift case is left out in these graphs since the
errors/prices are so big that they dominate all other methods.

The second graph in the top right of Figure \ref{fig:PFRA} shows prices for a
discretization grid of 1 step per tenor period. Here the prices for Picard and
Euler clearly reflect the higher discretization error and the PC and IPC methods
are indistinguishable, and significantly lower than Picard and Euler. Note that
prices are around the same level as in the 5 step per tenor period
discretization.

Finally, we price each FRA using a single long discretization step from time
zero to the maturity of each contract, also referred to as ``long-stepping''. In
this case the Picard and Euler methods are analogous to the frozen drift
approximation, while PC and Picard+PC are equivalent to each other; hence we
draw only PC, IPC, and frozen drift. We see, somewhat surprisingly that the
frozen drift slightly outperforms PC and IPC which contradicts the results in
the log-normal case previously studied in the literature. Nevertheless the
errors are quite high and beyond an acceptable level for all methods.

The general conclusion one can draw from these graphs is that Picard+PC with 1
discretization step per tenor period would be the preferable choice. The errors
are indistinguishable from regular PC and IPC, but the method has the advantage
that prices can be parallelized in the maturity dimension, and the gains in
computational time shown in the previous section can be realized.

\section{Conclusion}

This paper derives new approximation methods for Monte Carlo simulation in LIBOR
models. The methods address the problem of speed in Monte Carlo simulations by
allowing for parallel computing through Picard approximations. In particular,
our method decouples the interdependence of the rates when moving them forward
in time in a simulation, meaning that computations can be parallelized in the
maturity dimension. Furthermore, the largest computational load in the model
stems from the exponential growth of the drift terms. We reduce this growth to
quadratic through truncated expansions of the product terms that appear in the
drift. The accuracy is very high if the second order expansion is employed and
we showed that it reduces the computational load immensely.

Numerical methods for \lib market models driven by general semimartingales are
still in their infancy. As we have demonstrated there is still work to be done
in this area, in particular developing algorithms for pricing derivatives using
long time-stepping. Predictor-corrector methods do not perform very well in this
case and a finer discretization grid needs to be employed.

\bibliographystyle{chicago}
\bibliography{references}

\end{document}